\documentclass[10pt]{llncs}
\pdfoutput=1

\usepackage{epsfig}
\usepackage{color}
\usepackage{graphicx}
\usepackage{amsmath}
\usepackage{amssymb}

\long\def\symbolfootnote[#1]#2{\begingroup%
 \def\thefootnote{\fnsymbol{footnote}}\footnote[#1]{#2}\endgroup}

\newcommand{\ket}[1]{{|{#1} \rangle}}

\newcommand{\xor}{\ensuremath{\oplus}}

\newcommand{\transform}[1]{\stackrel{#1}{\mapsto}} 

\newcommand{\fanout}[1]{{F_{#1}}}
\newcommand{\zfanout}[1]{{Z_{#1}}}
\newcommand{\toffoli}[1]{{\wedge_{#1}(X)}}
\newcommand{\CNOT}{\ensuremath{\mathrm{CNOT}}}

\spnewtheorem{mydef}[theorem]{Definition}{\bfseries}{\rm}

\renewcommand{\marginpar}[1]{}

\begin{document}

\title{Universal Quantum Circuits}

\author{
Debajyoti Bera\inst{1}
\and
Stephen Fenner\inst{2}
\and
Frederic Green\inst{3}
\and
Steve Homer\inst{1}
}

\institute{
Boston University, Department of Computer Science, Boston, MA
02134.
\thanks{\email{$\{(\mathtt{dbera}|\mathtt{homer})\}$@cs.bu.edu}.
Partially supported by the National Security Agency (NSA)
and Advanced Research and Development Agency (ARDA) under Army Research
Office (ARO) contract number DAAD~19-02-1-0058.}
\and
University of South Carolina, Department of Computer Science
and Engineering, Columbia, SC 29208.
\thanks{\email{fenner@cse.sc.edu}. Partially supported by NSF grant CCF-05-15269.}
\and
Clark University, Department of Mathematics and Computer Science, Worcester,
MA 01610.
\thanks{\email{fgreen@black.clarku.edu}.
Partially supported by the NSA and ARDA under ARO contract number DAAD~19-02-1-0058.}
}

\maketitle

\begin{abstract}
We define and construct efficient depth-universal and almost-size-universal
quantum circuits.  Such circuits can be viewed as general-purpose simulators for
central classes of quantum circuits and can be used to capture the
computational power of the circuit class being simulated.  For depth we
construct universal circuits whose depth is the same order as the circuits
being simulated. For size, there is a log factor blow-up in the universal
circuits constructed here. We prove that this construction is nearly optimal.
\end{abstract}

\section{Introduction}

Like resource-bounded universal Turing machines, efficiently constructed
universal circuits capture the hardness of languages computed by circuits in a
given circuit class. As a result, the study of the existence and complexity of
universal circuits for quantum circuit classes provides insight into the
computational strength of such circuits, as well as their limits.

There is both a theoretical and a practical aspect to this study.  The
existence of a universal circuit family for a complexity class defined by
resource bounds (depth, size, gate width, etc.) provides an upper bound on the
resources needed to compute any circuit in that class. It also opens up
possibilities for proving lower bounds on the hard languages in the class, as
such bounds would follow from a lower bound proof for the language computed by
a universal circuit family for the circuit class.

More precisely, the specific, efficient construction of a universal circuit for
a class of circuits yields, for a fixed input size, a single circuit which can
be used to carry out the computation of every circuit (with that same input
size) in that family, basically a chip or processor for that class of
circuits.  The more efficient the construction of the universal circuit, the
smaller the processor for that class.

Furthermore, the universal circuit is in a sense a compiler for all possible
computations of all circuits in this family.  It can be used to efficiently
program all possible computations capable of being carried out by circuits in
this circuit class, and in doing so automatically acts as a general purpose
simulator and with as little loss of efficiency as is possible.

In the case of quantum circuits there are particular issues relating to the
requirements that computations must be clean and reversible which come into
play, and to an extent complicate the classical methods.  Still much of our
motivation for this work originates with classical results due to Cook,
Valiant, and others \cite{cook-hoover,val76}.
\marginpar {Details of their work in the next section. Is
that OK?}
Cook and Hoover considered depth universality and described a depth-universal uniform circuit family for circuits of depth $\Omega(\log n)$.
Valiant studied size universality and showed how to construct universal
circuits of size $O(s \log s)$ to simulate any circuit of size $s$.  (See Section~\ref{sec:other-work}.)




\begin{mydef}[Universal Quantum Circuits]
Fix $n>0$ and let $\mathcal{C}$ be a collection of quantum circuits on $n$ qubits.  A quantum circuit $U$ on $n+m$ qubits is \emph{universal for $\mathcal{C}$} if, for every circuit $C\in\mathcal{C}$, there is a string $x\in\{0,1\}^m$ (the \emph{encoding}) such that for all strings $y\in\{0,1\}^n$ (the \emph{data}),
\[ U(\ket{y}\otimes\ket{x}) = C\ket{y}\otimes \ket{x}. \]
\end{mydef}

The circuit collections we are interested in are usually defined by bounding various parameters such as the size (number of gates), depth (number of layers of gates acting simultaneously on disjoint sets of qubits), or palette of allowed gates (e.g., Hadamard, $\pi/8$, CNOT).

As in the classical case, we also want our universal circuits to be \emph{efficient} in various ways.  For one, we restrict them to using the same gate family as the circuits they simulate.  We may also want to restrict their size or the number $m$ of qubits they use for the encoding.  We are particularly concerned with the depth of universal circuits.

\begin{mydef}[Depth-Universal Quantum Circuits]
Fix a family $\mathcal{F}$ of unitary quantum gates.  A family of quantum circuits $\{U_{n,d}\}_{n,d>0}$ is \emph{depth-universal over $\mathcal{F}$} if
\begin{enumerate}
\item
$U_{n,d}$ is universal for $n$-qubit circuits with depth $\le d$ using gates from $\mathcal{F}$,
\item
$U_{n,d}$ only uses gates drawn from $\mathcal{F}$,
\item
$U_{n,d}$ has depth $O(d)$, and
\item
the number of encoding qubits of $U_{n,d}$ is polynomial in $n$ and $d$.
\end{enumerate}
\end{mydef}

Depth-universal circuits are desirable because they can simulate any circuit within a constant slow-down factor.  Thus they are as time-efficient as possible.

Our first result, presented in Section~\ref{sec:depth-univ}, shows that depth-universal quantum circuits exist for the gate families $\mathcal{F} = \{H,T\} \cup \{\fanout n \mid n\ge 1\}$ and $\mathcal{F}' = \{H,T\} \cup \{\fanout n \mid n\ge 1\} \cup \{\toffoli n \mid n\ge 1\}$, where $H$ and $T$ are the Hadamard and $\pi/8$ gates, respectively, and $\fanout n$ and $\toffoli n$ are the $(n+1)$-qubit fanout and $(n+1)$-qubit Toffoli gates, respectively (see Section~\ref{sec:prelims}).

\begin{theorem}\label{thm:depth-univ}
Depth-universal quantum circuits exist over $\mathcal{F}$ and over $\mathcal{F'}$.  Such circuits use $O(n^2d)$ qubits and can be built log-space uniformly in $n$ and $d$.
\end{theorem}

Note that the results for the two circuit families are independent, because it is not known whether $n$-qubit Toffoli gates can be implemented exactly in constant depth using single-qubit gates and fanout gates, although they can be approximated this way \cite{HS:fanout}.

It would be nice to find depth-universal circuits over families of bounded-width gates\footnote{The width of a gate is the number of qubits it acts upon.} such as $\{H,T,\CNOT\}$.  Depth-universal circuits with bounded-width gates, if they exist, must have depth $\Omega(\log n)$ and thus can only depth-efficiently simulate circuits with depth $\Omega(\log n)$.  This can be easily seen as follows: Suppose all you wanted was a universal circuit $U$ for depth-1 circuits on $n$ qubits that use CNOT gates \emph{only}.  Since any pair of the $n$ qubits could potentially be connected with a CNOT gate, that pair must be connected somehow (indirectly perhaps) within the circuit $U$.  Thus any data input qubit can potentially affect any of the other $n-1$ data output qubits.  Since $U$ only has constant-width gates, the number of qubits affected by any given data input increases by only a constant factor per layer, and so $U$ must have $\Omega(\log n)$ layers.

One can therefore only hope to find depth-universal circuits for circuits of depth $\Omega(\log n)$ over bounded-width gates.  Although such circuits exist in the classical case (see below), we are unable to construct them in the quantum case (see Section~\ref{sec:open}).


\subsection{Other relevant work}
\label{sec:other-work}

The study of quantum circuit complexity was originated by
Yao~\cite{yao}.  The basic definitions and first results in this research area
can be found in Nielsen and Chuang~\cite{nielsen-chuang}. Most of the research on
universal quantum circuit classes deals with finding small, natural, universal sets of
gates which can be used in quantum circuits to efficiently simulate any
quantum computation. Our problem and point of view here is quite different.
We have the goal of constructing, for a natural class $C$ of quantum circuits, a
single family of quantum circuits which can efficiently simulate all circuits
on the class $C$. In this paper we consider classes $C$ which have significant
resource bounds (small or even constant depth, or fixed size) and ask that the
corresponding universal circuits family to have similar depth or size bounds.

Cook and Hoover~\cite{cook-hoover} considered the problem of
constructing general-purpose classical (Boolean) circuits using gates with fanin two.  They asked whether, given
$n,c,d$, there is a circuit $U$ of size $c^{O(1)}$ and depth $O(d)$ that can
simulate any $n$-input circuit of size $c$ and depth $d$. Cook and Hoover constructed a
depth-universal circuit for depth $\Omega(\log n)$ and polynomial size, but which
takes as input a nonstandard encoding of the circuit, and they also presented a
circuit with depth $O(\log n \log \log n)$ to convert the standard encoding of
the circuit to the required encoding.

Valiant looked at a similar problem---trying to minimize the size of
the universal circuit~\cite{val76}. He considered classical circuits built
from fanin $2$ gates (but with unbounded fanout) and embedded the circuit in a
larger universal graph. Using switches at key vertices of the universal graph,
any graph (circuit) can be embedded in it. He managed to create universal
graphs for different types of circuits and showed how to construct a $O(c \log
c)$-size and $O(c)$-depth universal circuit.  He also showed that his
constructions have size within a constant multiplicative factor of
the information theoretic lower bound.

\marginpar{S. Fenner: Wasn't there some similar simple lower bound
for our depth-universal result?}
For quantum circuits, Nielsen and Chuang (in
\cite{nielsen-chuang-paper}) considered the problem of building generic
universal circuits, or \emph{programmable universal gate arrays} as they call them.
Their universal circuits work on two quantum registers, a data register and a
program register. They do not consider any size or depth bound on the circuits
and show that simulating every possible unitary operation requires completely
orthogonal programs in the program register.  Since there are infinitely many
possible unitary operations, any universal circuit would require an infinite
number of qubits in the program register.  This shows that it is not possible
to have a generic universal circuit which works for all circuits of a certain
input length.  However they showed that it is possible to construct an extremely weak type of probabilistic universal circuit with size linear in the number of inputs to
the simulated circuit.

Sousa and Ramos considered a similar problem of creating a
universal quantum circuit to simulate any quantum gate~\cite{sousa07}. They
construct a basic building block which can be used to implement any single-qubit or CNOT gate on $n$ qubits by switching certain gates on and off. They
showed how to combine several of these building blocks to implement any $n$-qubit quantum gate.

\subsection{Outline of the paper}


For the rest of the paper, we will use $U$ to denote the universal circuit and $C$ to denote the circuit being simulated.
We define the quantum gates we will use in Section~\ref{sec:prelims}.  The construction of depth-universal circuits is in Section~\ref{sec:depth-univ}.  We briefly describe the construction of almost-size-universal quantum circuits in Section~\ref{sec:size-univ}.  We mention a couple of miscellaneous results in Section~\ref{sec:misc}.


\section{Preliminaries}
\label{sec:prelims}

%


We assume the standard notions of quantum states, quantum circuits, and quantum gates described in \cite{nielsen-chuang}, in particular, $H$ (Hadamard), $T$ ($\pi/8$), $S = T^2$ (phase), and $\CNOT$ (controlled NOT).  We will also need some additional gates, which we now motivate.

The depth-universal circuits we construct require the
ability to feed the output of a single gate to many other gates. While this
operation, commonly known as fanout, is common in classical circuits, copying
an arbitrary quantum state unitarily is not possible in quantum circuits due to
the no-cloning theorem \cite{nielsen-chuang}.  It turns out that we can
construct our circuits using a classical notion of fanout operation,
defined as the \emph{fanout gate} $\fanout n : \ket{c,t_1,\ldots,t_n}
\mapsto \ket{c,c \xor t_1, \ldots, c \xor t_n}$ for any of the standard basis
states $\ket{c}$ (the control) and $\ket{t_1}, \ldots, \ket{t_n}$ (the targets) and extended linearly to other
states\footnote{This does not contradict the no-cloning theorem as only
classical states are copied.} \cite{ffghz}.  $\fanout n$ can be
constructed in depth $\lg n$ using $\CNOT$ gates. We need to use unbounded
fanout gates to achieve \marginpar{- explain {\it full?}}full depth
universality. We also use the \emph{unbounded Toffoli gate} $\toffoli n : \ket{c_1,\ldots,c_n,t} \mapsto \ket{c_1,\ldots,c_n,
t\oplus \bigwedge_{i=1}^n c_i}$.\marginpar{Our definition of Toffoli.}  We reserve the term ``Toffoli gate'' to refer to the (standard) Toffoli gate $\toffoli 2$, which is defined on three qubits.

In addition to the fanout gate, our construction requires us to use controlled versions of the gates used in the simulated circuit.  For
most of the commonly used basis sets of gates (e.g., Toffoli gate, Hadamard gate,
and phase gate $S$), the gates themselves are sufficient to construct their
controlled versions (e.g., a controlled Hadamard gate can be constructed using a
Toffoli gate and Hadamard and phase gates).  Depth or size universality requires that the controlled versions of the gates should be constructible
using the gates themselves within proper depth or size, as required.

\begin{mydef}[Closed under controlled operation]
A set of quantum gates $G=\{G_1, \ldots\}$ is said to be \emph{closed under
controlled operation} if for each $G_i \in G$, the controlled version of the
gate $\textrm{C-}G_i \ket{c}\ket{t} \longrightarrow \ket{c}G_i^c\ket{t}$ can
be implemented in constant depth and size using the gates in $G$. Here, $\ket{c}$ is a single qubit and $G_i$
could be a single or a multi-qubit gate.
\end{mydef}


Note that $\CNOT = \fanout 1$, and given $H$, $T$, and $\CNOT$ we can implement the Toffoli gate via a standard constant-size circuit \cite{nielsen-chuang}.  We can implement the phase gate $S$ as $T^2$, and since $T^8 = I$, we can implement $S^\dagger = T^6$ and $T^\dagger = T^7$.

A \emph{generalized $Z$ gate}, which we will hereafter refer to simply as a \emph{$Z$ gate}, is an extension of the single-qubit Pauli $Z$ gate ($\ket{x} \mapsto (-1)^x\ket{x}$) to multiple qubits:
\[ \ket{x_1, \cdots, x_n}
\transform{Z} (-1)^{x_1x_2\cdots x_n}\ket{x_1, \ldots, x_n}. \]
A $Z$ gate can
be constructed easily (in constant depth and size) from a single unbounded Toffoli gate (and vice versa) by conjugating the target qubit of the unbounded Toffoli gate with $H$ gates (i.e., placing $H$ on both sides of the Toffoli gate on its target qubit).

Similarly, a \emph{$Z$-fanout gate} $\zfanout n$ applies the single-qubit $Z$ gate to each of $n$ target qubits if the control qubit is set:
\marginpar{See footnote!}
\[ \ket{c, t_1, \cdots, t_n}
\transform{\zfanout n} (-1)^{c \cdot (t_1 + \cdots + t_n)}\ket{c, t_1, \ldots, t_n}. \]
A $\zfanout n$ gate can be constructed from a single $\fanout n$ gate and vice versa in constant depth (although not constant size) by conjugating each target with $H$ gates.  So, in our depth-universal circuit construction, we can use either or both of these types of gates.  Similarly for unbounded Toffoli versus $Z$ gates.  $Z$ gates and $Z$-fanout gates are important because they only change the phase, leaving the values of the qubits intact (they are represented by diagonal matrices in the computational basis).  This allows us to use a trick due to H{\o}yer and \v{S}palek \cite{HS:fanout} and run all possible gates for a layer in parallel.

\section{Depth-universal quantum circuits}
\label{sec:depth-univ}


In this section, we prove Theorem~\ref{thm:depth-univ}, i.e., that depth-universal circuits exist for each of the gate families
\begin{eqnarray*}
\mathcal{F} & = & \{H,T\} \cup \{\fanout n \mid n\ge 1\}, \\
\mathcal{F}' & = & \{H,T\} \cup \{\fanout n \mid n\ge 1\} \cup \{\toffoli n \mid n\ge 1\}.
\end{eqnarray*}
We first give the proof for $\mathcal{F}$ then show how to modify it for $\mathcal{F}'$.

The depth-universal circuit $U$ we construct simulates the input circuit $C$ layer by layer, where a layer
consists of the collection of all its gates at a fixed depth.  $C$ is encoded in a slightly altered form, however.  First, all the fanout gates in $C$ are replaced with $Z$-fanout gates on the same qubits with $H$ gates conjugating the targets.  At worst, this may roughly double the depth of $C$ (adjacent $H$ gates cancel).  Each layer of the resulting circuit is then separated into three adjacent layers: the first having only the $H$ gates of the original layer, the second only the $T$ gates, and the third only the $Z$-fanout gates.
$U$ then simulates each layer of the modified $C$ by a constant number of its own layers.  We describe next how these
layers are constructed.

\begin{figure}[ht]
\begin{center}\resizebox{.875\textwidth}{!}{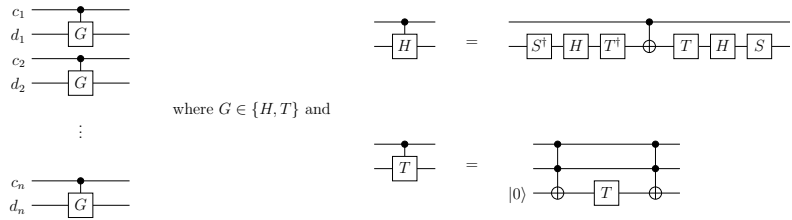}
\caption{Simulating a layer of single-qubit $G$ gates with controlled $G$ gates.  The ancilla in the implementation of the controlled $T$ gate is assumed part of the encoding.  The ancilla is
reset to $0$ at the end and hence can be reused for implementing all $T$ layers.}
\label{fig:single-qubit}
\end{center}
\end{figure}

\paragraph{Simulating single-qubit gates.} The circuit to simulate an $n$-qubit layer of
single-qubit gates of type $G$, say, consists of a layer of controlled-$G$
gates where the control qubits are fed from the encoding and the target qubits
are the data qubits.  Figure~\ref{fig:single-qubit} shows a layer of $G$
gates, where $G\in\{H,T\}$, controlled using $H$, $S$, $T$, CNOT, and Toffoli gates.
\marginpar{How to do other controlled single qubit gates? Do we need more
examples? We already show our gates are closed under control.} To simulate
$G$ gates on qubits $i_1, \ldots, i_k$, say, set $c_{i_1}, \ldots,
c_{i_k}$ to $1$ and the rest of the $c$-qubits to $0$.

\begin{figure}[ht]
\begin{center}
\resizebox{!}{.35\textheight}{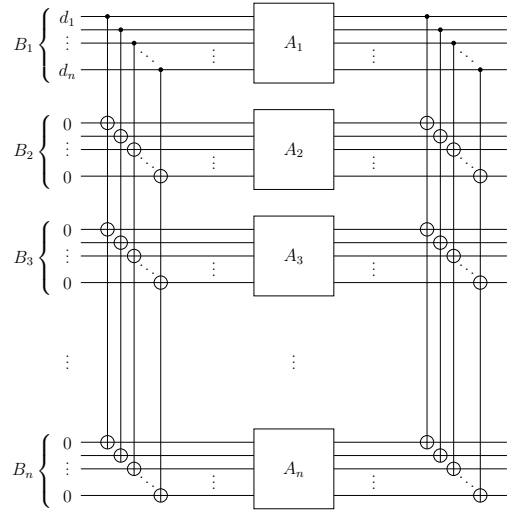}
\caption{Simulating a layer of $Z$-fanout gates.}
\label{fig:Z-fanout}
\end{center}
\end{figure}

\paragraph{Simulating $Z$-fanout gates.} The circuit to simulate a
$Z$-fanout layer is shown in Figure~\ref{fig:Z-fanout}.
The top $n$ qubits are the original data qubits.  The rest are ancilla qubits. 
All the qubits are arranged in $n$ blocks $B_1,\ldots,B_n$ of $n$ qubits per
block.  The qubits in block $B_i$ are labeled $b_{i1},\ldots,b_{in}$.

\begin{figure}[h]
\begin{minipage}{0.45\linewidth}
\centering
\resizebox{!}{\linewidth}{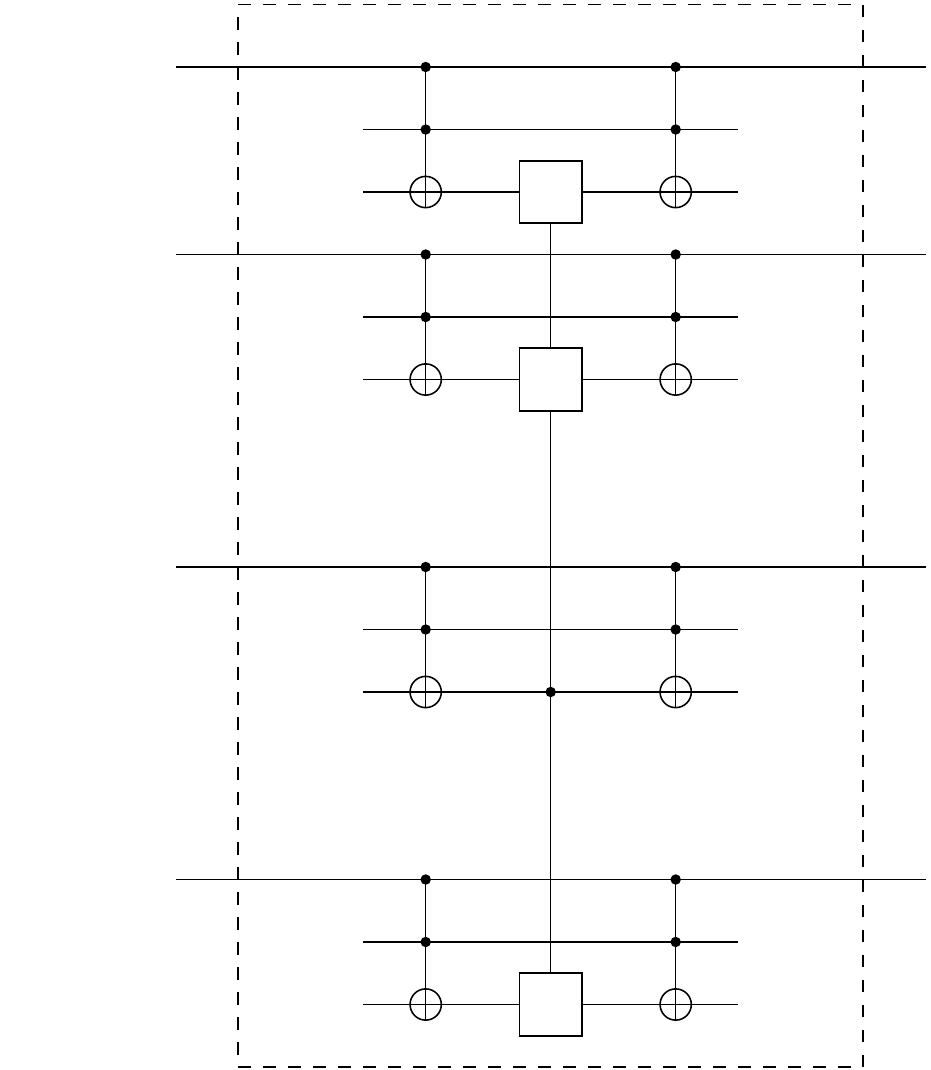}
\caption{Subcircuit $A_i$ in the simulation of $Z$-fanout gates.}
\label{fig:A-i-Z-fanout}
\end{minipage}
\hspace{0.25in}
\begin{minipage}{0.45\linewidth}
\centering
\resizebox{!}{\linewidth}{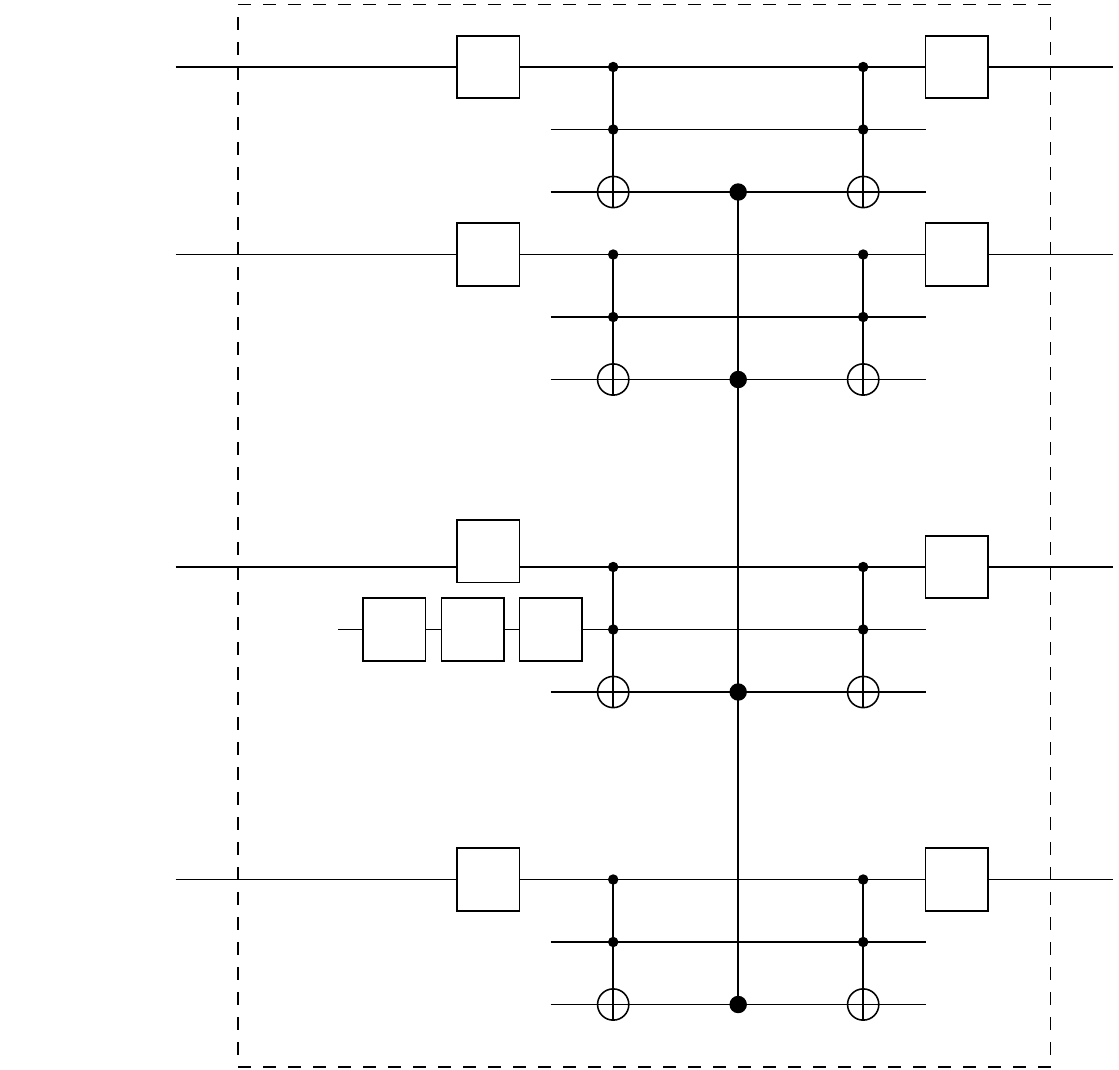}
\caption{Subcircuit $A_i$ for a layer of $Z$ gates.}
\label{fig:A-i-Z}
\end{minipage}
\end{figure}

Each $A_i$ subcircuit looks like Figure~\ref{fig:A-i-Z-fanout}.  The
qubits $c_{i1},\ldots,c_{in}$ are encoding qubits.
The large gate between the two columns of Toffoli gates is a $Z$-fanout gate with its
control on the $i$th ancilla (corresponding to $b_{ii}$ and $c_{ii}$) and targets on all the other ancill\ae.

Here is the state evolution from $\ket{\vec{d}} = \ket{d_1\cdots d_n}$, suppressing the $c_{ij}$ qubits and ancill\ae\ internal to the $A_i$ subcircuits
in the ket labels.  Note that after the first layer of fanouts, each qubit
$b_{ij}$ carries the value $d_j$.
\begin{eqnarray*}
\ket{\vec{d},\vec{0},\ldots,\vec{0}} & \mapsto &
\ket{\vec{d},\vec{d},\ldots,\vec{d}} \\
& \mapsto & (-1)^{\sum_i d_ic_{ii}\left(\sum_{j\ne i}
d_jc_{ij}\right)}\ket{\vec{d},\vec{d},\ldots,\vec{d}} \\
& \mapsto & (-1)^{\sum_i d_ic_{ii}\left(\sum_{j\ne i}
d_jc_{ij}\right)}\ket{\vec{d},\vec{0},\ldots,\vec{0}}.
\end{eqnarray*}

To simulate some $Z$-fanout gate $G$ of $C$ whose control is on the $i$th qubit, say,
we do this in block $B_i$ by setting $c_{ii}$ to $1$ and setting $c_{ij}$ to $1$
for every $j$ where the $j$th qubit is a target of $G$.  All the other
$c$-qubits in $B_i$ are set to $0$.  We can do this in separate blocks for
multiple $Z$-fanout gates on the same layer, because no two gates can share the same
control qubit.  Any $c$-qubits in unused blocks are set to $0$.

\paragraph{Simulating unbounded Toffoli gates.}  We can modify the construction above to accommodate unbounded Toffoli gates (gate family $\mathcal{F}'$), or equivalently $Z$ gates, by breaking each layer of $C$ into four adjacent layers, the first three being as before, and the fourth containing only $Z$ gates.
The top-level circuit to simulate a layer of $Z$ gates looks just as before (Figure~\ref{fig:Z-fanout}), except now each $A_i$ subcircuit looks a bit different and is shown in Figure~\ref{fig:A-i-Z}, where the central gate is a $Z$ gate connecting the ancill\ae.

As before, the qubits $c_{i1},\ldots,c_{in}$ are encoding qubits.
The $XZX$ gates on $c_{ii}$ multiply the overall phase by $(-1)^{\overline{c_ii}}$.
When the $Z$ gate of $A_i$ is applied, its $j$th contact point is in the state
$\overline{\overline{b_{ij}} c_{ij}}$. Note that
$\overline{\overline{b_{ij}} c_{ij}} = b_{ij}$ if $c_{ij}=1$ and $1$ otherwise.
The $Z$ gate then multiplies the overall phase by $(-1)^{\prod_j (\overline{\overline{b_{ij}} c_{ij}})} =
(-1)^{\prod_{j:c_{ij}=1} b_{ij}}$.
The state thus evolves as given below:
\begin{eqnarray*}
\ket{\vec{d},\vec{0},\ldots,\vec{0}} & \mapsto &
\ket{\vec{d},\vec{d},\ldots,\vec{d}} \\
& \mapsto & (-1)^{\sum_i \overline{c_{ii}} + \prod_{j:c_{ij}=1}
b_{ij}}\ket{\vec{d},\vec{d},\ldots,\vec{d}} \\
& \mapsto & (-1)^{\sum_i \overline{c_{ii}} + \prod_{j:c_{ij}=1}
b_{ij}}\ket{\vec{d},\vec{0},\ldots,\vec{0}}
\end{eqnarray*}
To simulate some $Z$ gate $G$ of $C$ whose first qubit is $i$, say, we do this in block
$B_i$ by setting $c_{ii}$ to $1$ and setting $c_{ij}$ to $1$ for every $j$ where
the $j$th qubit is part of $G$.  All the other $c$-qubits in $B_i$ are
set to $0$.  As before, we can do this in separate blocks for multiple gates on
the same layer, because no two gates can share the same first qubit.  Any
$c$-qubits in unused blocks are set to $0$, and it is easy to check that this makes the block have no net effect.

\section{Size-universal quantum circuits}
\label{sec:size-univ}


Similar to a depth-universal circuit, a
\emph{size-universal circuit} is a universal circuit with the same order of the
number of gates as the circuit it is simulating. Formally,\marginpar{Explain
universal circuit for fanin-2 gates}

\begin{mydef}
A family $\{U_{n,c}\}$ of universal circuits for $n$-qubit circuits of size $\le c$ is \emph{size-universal} if
$\textrm{SIZE}(U_{n,c}) = O(c)$.
\end{mydef}

A simple counting argument shows that it is not possible to obtain a
completely size-universal circuit for fanin-$2$ circuits. Consider all
circuits with $c$ fanin-$2$ gates where one input of each gate is the first
qubit. There are $(n-1)^c$ possible circuits. Then consider similar circuits
where there is no gate with input as the first qubit and continue recursively. Thus the
number of possible fanin-$2$ circuits is $\Omega((n-1)^{c+1})$. Since all the
encoding bits have to be connected to some of the fanin-$2$ gates in the
universal circuit, it must have $\Omega(c\log n)$ gates.

We use Valiant's idea of universal graphs \cite{val76} to construct a universal family of fanin-$2$ circuits that are very
close to the afforementioned lower bound. As before, we would like to simulate
$C$ by using the same set of gates used in $C$. Our construction works for any
circuit using unbounded Toffoli gates and any set of single-qubit and $2$-qubit gates
closed under the controlled operation.


First we will define a universal directed acyclic graph with $n$ special
vertices (called \emph{poles}) in which we can embed any circuit with $n$ gates
(considering the inputs also as gates). The embedding will map the wires in the
circuit to paths in the graph.

\begin{mydef}[Edge-embedding \cite{val76}]
An \emph{edge-embedding} $\rho$ of $G=(V,E)$ into $G'=(V',E')$ maps $V$ one-to-one to
$V'$ and maps each edge $(i,j)\in E$ to a directed path $\rho(i)\leadsto
\rho(j)$ in $G'$ such that distinct edges are mapped to edge-disjoint paths.
\end{mydef}

The graph of any circuit of size $n$ can be represented as a directed acyclic
graph with vertices $\{1, \ldots, n\}$ such that there is no edge from $j$ to
$i$ for $i<j$ and each vertex has fanin and fanout $2$. Let $\Gamma_2(n)$ be
the set of all such graphs.

\begin{mydef}[Edge-universal graph \cite{val76}]
A graph $G'$ is \emph{edge-uni\-versal} for $\Gamma_2(n)$ if it has distinct poles $p_1,\ldots,p_n$ such that
any graph $G\in\Gamma_2(n)$ can be edge-embedded into $G'$ where each vertex
$i \in G$ is mapped to vertex $\rho(i)=p_i \in G'$.
\end{mydef}

Then, Valiant shows how to construct a universal graph.

\begin{theorem}[\cite{val76}]
There is a constant $k$ such that for all $n$ there exists an acyclic graph $G'$ that is edge-universal
for $\Gamma_2(n)$, and $G'$ has $kn\lg n$ vertices, each vertex
having fanin and fanout $2$.
\end{theorem}

It is fairly easy to construct a universal circuit using the universal graph. In
fact, the universal circuit for circuits with $n$ inputs and $c$ gates will be any edge-universal graph
for $\Gamma_2(n+c)$.

\begin{figure}[h]
\begin{minipage}{0.45\linewidth}
\centering
\resizebox{1.15\linewidth}{!}{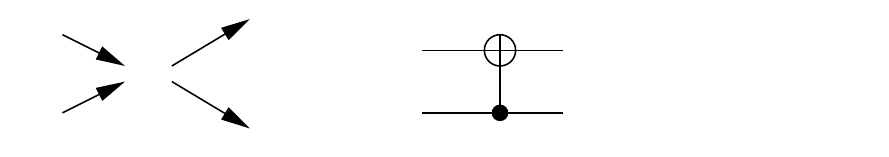}
\caption{The gate for a pole vertex $p_i$ is mapped to input $x_i$.}
\label{fig:size-univ-input-gate}
\end{minipage}
\hspace{0.25in}
\begin{minipage}{0.45\linewidth}
\centering
\resizebox{0.625\linewidth}{!}{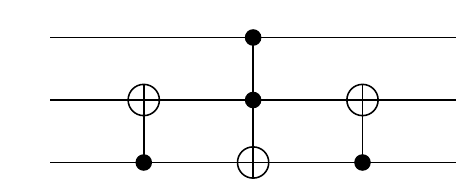}
\caption{The gates at a non-pole vertex $v$. The encoding bit $c_v$ specifies if
first output qubit should be mapped to first input or second input qubit and
similarly for second output qubit.}
\label{fig:size-univ-non-pole}
\end{minipage}
\end{figure}

\begin{figure}[h]
\centering
\resizebox{0.5\linewidth}{!}{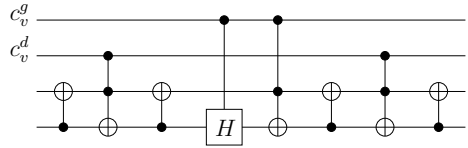}
\caption{Example of the gates at a pole vertex $v$ simulating a circuit
with $\CNOT$ and $H$ gates. The encoding bits $c_v^g$ specify which kind of gate
is at vertex $v$, and the $c_v^d$ specify which qubit the gate acts on (for $H$
gate) or which is the control qubit (for $\CNOT$ gate).}
\label{fig:size-univ-pole}
\end{figure}

Consider any such edge-universal graph $G'$. Then $G'$ has $c'=k(n+c)\log (n+c)$ vertices for some $k$. These $c'$ vertices
include fixed poles $p_1, \ldots, p_n, p_{n+1}, \ldots, p_{n+c}$ and
non-pole vertices. Create a quantum circuit $C'$ with $c'$ gates (including the
inputs and outputs) where $G'$ describes how
the gates connect to each other. For each of the vertices $p_1, \ldots, p_n$ of
$G'$, remove their incoming edges and replace the vertices by the input
as shown in Figure~\ref{fig:size-univ-input-gate}. Replace each of the vertices
$p_{n+1}, \ldots, p_{n+c}$ with a subcircuit that applies any of
the single- or $2$-qubit gates on the inputs, where the gate to apply is controlled by the
encoding. E.g., Figure~\ref{fig:size-univ-pole} shows the gates at a pole
vertex in a universal circuit simulating $\CNOT$ and $H$ gates. For a
non-pole vertex, replace it with a subcircuit that swaps the incoming and
outgoing wires (i.e., first input is connected to second output and second
input is connected to first output) or directly connects them (i.e., first input
is connected to first output and similarly for the second input). Again, the
subcircuit is controlled by the encoding which controls whether to swap or
directly connect (see Figure~\ref{fig:size-univ-non-pole}). The edge disjointness
property guarantees that wires in the embedded circuit are mapped to paths in
$C'$ which can share a vertex but cannot share any edge.


To simulate any fanin-$2$ circuit $C$ with $c$ gates acting on $n$ qubits, construct the
edge-universal graph $G'$ for $\Gamma_2(n+c)$.
Embed the graph of $C$ into $G'$ such that the input nodes of $C$ are
mapped to the poles $p_1, \ldots, p_n$ in $G'$. Now for each gate of the circuit,
consider the pole to which it was mapped.  Set a bit in the encoding to denote the type of
the gate at that pole.  For the non-pole vertices, set a bit in the encoding to
specify whether the two input values should be swapped or mapped directly to the
two output values. The size of the encoding is
$(n+c)(\log|\mbox{gates}|+1)+(|\Gamma_2(n+c)|-(n+c))$ which is $O(c \log c)$
for polynomial-size circuits.
This construction gives us a universal circuit with a logarithmic blow-up in
size.

\begin{theorem}
There is a constant $k$ and a family of universal circuits $U_{n,c}$ that can
simulate every circuit with $c$ gates acting on $n$ qubits such that
$\textrm{SIZE}(U_{n,c})=k(n+c)\log (n+c)$.
\end{theorem}

We can use a similar idea for circuits with unbounded fanin. First we decompose the unbounded fanin gates
using bounded fanin gates (fanin $2$ in this case). This is doable for most of
the common unbounded fanin gates. For example, an unbounded Toffoli gate of
size $f$ can be constructed using $\Theta(f)$ successive Toffoli gates of size
$3$, which can in
turn be implemented using Hadamard, phase, $\pi/8$ and $\CNOT$ gates~\cite{nielsen-chuang}. So any circuit of size $c$ consisting of Hadamard,
$\pi/8$ and unbounded Toffoli gates can be transformed into an equivalent
circuit with size at most $O(cn)$ consisting of these single-qubit gates and $\CNOT$ gates. The rest of the construction follows as before.

\begin{corollary}
There is a family of universal circuits $U_{n,c}$ that can simulate quantum
circuits of size $c$ on $n$ qubits and consisting of Hadamard, $\pi/8$, and
unbounded Toffoli gates such that $\textrm{SIZE}(U_{n,c}) = O(nc\log(nc))$.
\end{corollary}

\section{Other results}
\label{sec:misc}

\paragraph{Circuit encoding.}
We have been mostly concerned with the actual simulation of a quantum circuit $C$ by the universal
circuit $U$.  It is possible, however, to hide some complexity of the simulation in
$U$'s description of $C$ itself. Usually, the description of a classical circuit
describes the underlying graph of the circuit and specifies the gates at each
vertex. We can similarly describe a quantum circuit by its graph structure.
The description is extremely compact with size proportional to the size of the
circuit.
However, we use a description that is more natural for quantum circuits and
especially suitable for simulation. The description stores the grid structure
of the circuit; the rows of the grid correspond to the qubits, and the columns
correspond to the different layers of the circuit. This description is not
unique for any given circuit and its size is $O(nd)$, where $n$ is the number
of qubits and $d$ is the depth of the circuit. A graph-based description can
be easily converted to this grid-based description in polynomial time.

\paragraph{Depth-universal classical circuits.}
The techniques of Section~\ref{sec:depth-univ} can be easily adapted to build depth-universal circuits for a variety of classical (Boolean) circuit classes with unbounded gates, e.g., AC, ACC, and TC circuits.  The key reason is that these big gates are all ``self-similar'' in the sense that fixing some of the inputs can yield a smaller gate of the same type.  We will present these results in the full paper.

\section{Open Problems}
\label{sec:open}

A number of natural, interesting open problems
remain.

Fanout gates are used in our construction of a depth-universal
circuit family.  Is the fanout gate necessary in our construction?  We
believe it is.  In fact, we do not know how to simulate depth-$d$ circuits over $\{H,T,\CNOT\}$ universally in depth $O(d)$ without using fanout gates, even assuming that the circuits being simulated have depth $\Omega(\log n)$.
The shallowest universal circuits with bounded-width gates we know of have a $\lg n$ blow-up factor in the depth, just by replacing the fanout gates with log-depth circuits of $\CNOT$ gates.


Our results apply to circuits with very specific gate sets. How much can these
gate sets be generalized? Are similar results possible for any countable set of gates
containing Hadamard, unbounded Toffoli, and fanout gates?


We showed how to contruct a universal circuit with a logarithmic blow-up in
size. The construction is within a constant factor of the minimum possible
size for polynomial-size, bounded-fanin circuits. However for constant-size
circuits, we believe the lower bound can be tightened to match the proven
upper bound. For unbounded-fanin circuits, we construct a universal circuit
with size $O(nc \log nc)$ which is significantly larger than the bounded fanin
lower bound of $\Omega(c \log n)$. We think that a better lower bound is possible for the unbounded-fanin case.

\section*{Acknowledgments}

We thank Michele Mosca and Debbie Leung for insightful discussions.  The second author is grateful to Richard Cleve and IQC (Waterloo) and to Harry Buhrman and CWI (Amsterdam) for their hospitality.


\begin{thebibliography}{[FFGHZ06]}

\bibitem[BGH07]{bgh-sigact}
D.~Bera, F.~Green and S.~Homer.
\newblock Small depth quantum circuits.
\newblock {\em SIGACT News}, 38(2):35--50, 2007.

\bibitem[CH85]{cook-hoover}
Stephen~A.~Cook and H.~James~Hoover.
\newblock A depth-universal circuit.
\newblock {\em SIAM Journal on Computing}, 14(4):833--839, 1985.

\bibitem[FFGHZ06]{ffghz}
M.~Fang, S.~Fenner, F.~Green, S.~Homer and Y.~Zhang.
\newblock Quantum lower bounds for fanout.
\newblock {\em Quantum Information and Computation}, 6(1):046--057, 2006.

\bibitem[HS05]{HS:fanout}
P.~H{\o}yer and R.~{\v{S}}palek.
\newblock Quantum circuits with unbounded fan-out.
\newblock {\em Theory of Computing}, 1:81--103, 2005.

\bibitem[NC97]{nielsen-chuang-paper}
M.~A.~Nielsen and I.~L.~Chuang.
\newblock Programmable quantum gate arrays.
\newblock {\em Phys. Rev. Lett.}, 79(2):321-324, 1997.

\bibitem[NC00]{nielsen-chuang}
M.~A.~Nielsen and I.~L.~Chuang.
\newblock {\em Quantum Computation and Quantum Information}.
\newblock Cambridge University Press, 2000.

\bibitem[SR07]{sousa07}
P.~B.~M. Sousa and R.~V. Ramos.
\newblock Universal quantum circuit for $n$-qubit quantum gate: A programmable quantum gate.
\newblock {\em Quantum Information and Computation}, 7(3):228--242, 2007.

\bibitem[Val76]{val76}
Leslie G. Valiant.
\newblock Universal circuits (preliminary report).
\newblock In {\em Proceedings of the 8th ACM Symposium on the Theory of Computing}, 196--203, 1976.

\bibitem[Yao]{yao}
A.~C.-C. Yao.
\newblock Quantum circuit complexity.
\newblock In {\em Proceedings of the 34th IEEE Symposium on Foundations of Computer Science}, 352--361, 1993.

\end{thebibliography}
\end{document}